\documentclass[twocolumn,floatfix,preprintnumbers,citeautoscript,newabstract,nofootinbib]{revtex4} 
\usepackage{amsmath,amssymb,graphics}
\usepackage{bbm}

\newcommand{\E}[1]{\ensuremath{\mathrm{E}_{#1}}} 

\newcommand{\SO}[1]{\ensuremath{\mathrm{SO}(#1)}}
\newcommand{\SU}[1]{\ensuremath{\mathrm{SU}(#1)}}
\newcommand{\U}[1]{\ensuremath{\mathrm{U}(#1)}}
\newcommand{\Z}[1]{\ensuremath{\mathbbm{Z}_{#1}}} 
\DeclareMathOperator{\re}{Re}

\begin{document} \setlength{\unitlength}{1in}

\preprint{CERN-PH-TH/2006-238; NSF-KITP-06-112; \\}
\preprint{OHSTPY-HEP-T-06-007; TUM-HEP-652/06\\[0.5cm] }

\title{Low Energy Supersymmetry from the Heterotic Landscape\\[1cm]}

\author{{\bf\normalsize
Oleg~Lebedev$^{1}$\!,
Hans-Peter~Nilles$^2$\!,
Stuart~Raby$^3$\!, 
Sa\'ul~Ramos-S\'anchez$^2$\!,}\\{\bf\normalsize
Michael~Ratz$^4$\!,
Patrick~K.~S.~Vaudrevange$^2$\!,
Ak{\i}n~Wingerter$^3$}\\[0.5cm]
{\it\normalsize
${}^1$ CERN, Theory Division, CH-1211 Geneva 23, Switzerland}\\[0.15cm]
{\it\normalsize
${}^2$ Physikalisches Institut der Universit\"at Bonn,}\\[-0.05cm]
{\it\normalsize Nussallee 12, 53115 Bonn,
Germany}\\[0.15cm]
{\it\normalsize
${}^3$ Department of Physics, The Ohio State University,}\\[-0.05cm]
{\it\normalsize
191 W.\ Woodruff Ave., Columbus, OH 43210, USA}\\[0.15cm]
{\it\normalsize
${}^4$ Physik Department T30, Technische Universit\"at M\"unchen,}\\[-0.05cm]
{\it\normalsize James-Franck-Strasse, 85748 Garching,
Germany}
}

\begin{abstract}
We study possible correlations between properties  of the observable and hidden
sectors in  heterotic string theory. Specifically, we analyze  the case of the
\Z6-II   orbifold compactification which produces  a  significant  number of 
models with the spectrum of the supersymmetric standard model. We find that 
requiring realistic features does affect the hidden sector such that hidden
sector gauge group factors \SU4 and \SO8 are favoured. In the context of gaugino
condensation, this implies low energy supersymmetry breaking. 
\end{abstract}

\pacs{\dots}

\maketitle

In the string theory landscape \cite{Lerche:1986cx,Bousso:2000xa, Susskind:2003kw,Douglas:2003um,Douglas:2004qg}, the minimal
supersymmetric  standard model (MSSM) corresponds to a certain  subset of 
vacua  out of a huge variety.  To obtain string theory predictions, one can
first identify vacua with realistic properties, and then analyze their common
features. In this letter, we study possible implications of this approach for
supersymmetry breaking. First, we look for models consistent with the MSSM at
low energies, then we study common  features of their hidden sectors which are
responsible for  supersymmetry breaking.

We find that requiring realistic features affects the hidden sector such that,
in the context of gaugino condensation,   low energy supersymmetry breaking is
favoured. Since high energy supersymmetry is usually required by consistency of 
string models, this correlation provides a top--down motivation for low energy
supersymmetry, which is favoured by phenomenological considerations such as  the
gauge hierarchy problem and electroweak symmetry breaking.

We base our study  \cite{Lebedev:2006kn}    on the orbifold compactifications 
\cite{Dixon:1985jw,Dixon:1986jc}  of the  \E8$\times$\E8 heterotic string
\cite{Gross:1984dd}. 
Recent work on an orbifold GUT interpretation of heterotic models 
\cite{Kobayashi:2004ud,Forste:2004ie,Kobayashi:2004ya}
has facilitated construction of realistic models.
In particular,  the \Z6-II orbifold (see 
\cite{Kobayashi:2004ya}) 
has been shown to
produce many models with realistic features \cite{Buchmuller:2005jr,Buchmuller:2006ik,Lebedev:2006kn}.
These include the gauge group and
the matter content of the MSSM, gauge coupling unification and a heavy top
quark.  Such models are generated using the gauge shifts
\begin{eqnarray}
V^{ \SO{10},1}= &
\left(\tfrac{1}{3},\,\tfrac{1}{2},\,\tfrac{1}{2},\,0,\,0,\,0,\,0,\,0\right)&\left(\tfrac{1}{3},\,0,\,0,\,0,\,0,\,0,\,0,\,0\right) \;,
\nonumber \\ 
V^{ \SO{10},2 }= &
\left(\tfrac{1}{3},\,\tfrac{1}{3},\,\tfrac{1}{3},\,0,\,0,\,0,\,0,\,0\right)&\left(\tfrac{1}{6},\,\tfrac{1}{6},\,0,\,0,\,0,\,0,\,0,\,0\right) \;, \nonumber
\label{eq:so10shifts}
\end{eqnarray}
and 
\begin{eqnarray}
 V^{\E6 , 1}= &
\left(\tfrac{1}{2},\,\tfrac{1}{3},\,\tfrac{1}{6},\,0,\,0,\,0,\,0,\,0\right)&\left(0,\,0,\,0,\,0,\,0,\,0,\,0,\,0\right)\;,
\nonumber \\ 
 V^{ \E6 ,2}= &
\left(\tfrac{2}{3},\,\tfrac{1}{3},\,\tfrac{1}{3},\,0,\,0,\,0,\,0,\,0\right)&\left(\tfrac{1}{6},\,\tfrac{1}{6},\,0,\,0,\,0,\,0,\,0,\,0\right).  \nonumber  \label{eq:e6shifts}
\end{eqnarray}
These shifts are chosen due to their ``local grand unified theory (GUT)'' 
\cite{Buchmuller:2004hv,Buchmuller:2005jr,Buchmuller:2005sh,Buchmuller:2006ik}
properties. They lead to massless matter in the first twisted sector ($T_1$)
forming a  $\boldsymbol{16}$--plet of SO(10) in the case of $V^{ \SO{10},1}, V^{
\SO{10},2} $, and $\boldsymbol{27}$--plet of \E6 in the case of $V^{\E6 ,
1},V^{\E6 , 2}$. These states are invariant under the orbifold action and all
appear in  the low energy theory. Further,  if we choose Wilson lines such that 
\begin{equation}
 G_\mathrm{SM}~ \subset~ \SU5 \subset \SO{10} ~\text{or}~ \E6 \;,
\end{equation} 
the hypercharge will be  that of standard 4D GUTs. These features facilitate
construction of realistic models.

We focus on  models with one Wilson line of order 3 ($W_3$) and one Wilson line
of order 2 ($W_2$), although we include all models with 2 Wilson lines
in the statistics. These are the simplest constructions allowing for
3 MSSM matter families without chiral exotics. In this case, two matter
generations have similar properties while the third family is different.
Selection of realistic models proceeds as follows: 
\renewcommand{\labelenumi}{(\arabic{enumi})}
\begin{enumerate}
 \item Generate Wilson lines $W_3$ and $W_2$.
 \item Identify ``inequivalent'' models.
 \item Select models with $G_\mathrm{SM} \subset \SU5 \subset \SO{10}$.
 \item Select models with net three $(\boldsymbol{3},\boldsymbol{2})$.
 \item Select models with non--anomalous $\U1_{Y} \subset \SU5$.
 \item Select models with net 3 SM families + Higgses + vector--like.
  \item Select models with a heavy top. 
 \item Select models where exotics decouple and gauginos condense.
\end{enumerate}  
Steps (1)--(7)  are described in detail in Ref.~\cite{Lebedev:2006kn}.
At the last Step, we select models in which the decoupling of the SM exotic
states is possible without breaking the largest gauge group in the hidden
sector. We find that  all or almost all of the  matter states  charged  under
this group can be given large masses consistent with string selection rules,  
 which allows for  spontaneous supersymmetry
breaking via   gaugino condensation.

\begin{table*}[t!]
\centerline{
\begin{tabular}{l||l|l||l|l}
 criterion & $V^{\SO{10},1}$ & $V^{\SO{10},2}$ & $V^{\E6,1}$ & $V^{\E6,2}$\\
\hline
&&&&\\
  (2) inequiv. models with 2 WL
  &$22,000$ & $7,800$  &$680$  &$1,700$ \\[0.2cm]
  (3) SM gauge group $\subset$ SU(5) $\subset$ SO(10)
  (or \E6)
  &3563 &1163 &27 &63\\[0.2cm]
  (4) 3 net $(\boldsymbol{3},\boldsymbol{2})$
  &1170 &492 &3 &32\\[0.2cm]
  (5) non--anomalous $\U1_{Y}\subset \SU5 $
  &528 &234 &3 &22\\[0.2cm]
  (6) spectrum $=$ 3 generations $+$ vector-like
  &128 &90 &3 &2\\[0.2cm]
  (7) heavy top
  &72 &37 &3 &2\\[0.2cm]
  (8) exotics decouple +   gaugino condensation
  &47 & 25 & 3 & 2\\
\hline
\end{tabular}
}
\caption{Statistics of \Z6-II orbifolds based on the shifts
$V^{\SO{10},1},V^{\SO{10},2},V^{\E6,1},V^{\E6,2}$ with two Wilson lines. 
\label{tab:Summary} }
\end{table*}


The models satisfying all of the above criteria we consider  the ``MSSM
candidates''. Our results are presented in Table \ref{tab:Summary}.
More details can be found in \cite{WebTables:2006ml2}. 
We find it remarkable that out of ${\cal  O} (10^4)$ inequivalent 
models, ${\cal  O} (10^2)$ pass all of our requirements. In this sense, the
region of the heterotic landscape endowed with local  \SO{10} and \E6  GUTs
is particularly ``fertile'' \cite{Lebedev:2006kn}.

A comment is in order. 
We require that only the fields neutral under the SM and the largest
hidden sector group factor develop VEVs. 
 In ``generic'' vacua, the hidden sector gauge group is
broken by matter VEVs charged under this group. Similarly, the SM gauge group is
broken by generic vacuum configurations. Clearly, most of the string landscape
is not relevant to our physical world. It is only possible to obtain useful
predictions from the landscape once certain criteria are imposed. Here we
require that gaugino condensation be allowed so that supersymmetry can be
broken. Since the largest hidden sector group factor would dominate SUSY
breaking, we focus on vacua in which this factor is preserved by matter VEVs.
Within the set of  our promising models, we can now  study predictions for the
scale of supersymmetry breaking.

Our MSSM candidates  have the necessary ingredients for supersymmetry breaking
via gaugino condensation in the hidden sector
\cite{Nilles:1982ik,Ferrara:1982qs,Derendinger:1985kk,Dine:1985rz}.  In
particular, they contain non--Abelian gauge groups with little or no  matter.
The corresponding gauge interactions become strong at some  intermediate scale
which   can lead to spontaneous supersymmetry breakdown.  The specifics  depend
on the moduli stabilization mechanism, but the main features such as the scale
of supersymmetry breaking hold more generally. In particular, the gravitino mass
is related to the gaugino condensation scale 
$\Lambda \equiv \langle \lambda \lambda \rangle^{1/3} $  by
\begin{equation}
m_{3/2} \sim {\Lambda^3 \over M_\mathrm{Pl}^2} \;,
\end{equation}
while the proportionality constant is model--dependent. 
As an example, below we  consider  a well known mechanism based on
non--perturbative corrections to the K\"ahler potential.

The gaugino condensation scale 
is given by the renormalization group (RG)  invariant scale of the condensing gauge group,
\begin{equation}
 \Lambda~\sim~ 
 M_\mathrm{GUT}\,\exp \left(
 -\frac{1}{2\beta}\,\frac{1}{g^2(M_\mathrm{GUT})}
 \right) \;,
\label{Lambda}
\end{equation}
where $\beta$ is the beta--function. Since $1/g^2 =\re S$, this translates into
a superpotential for the dilaton $S$, $W\sim \exp(-3 S /2\beta )$. This simple
superpotential suffers from the notorious ``run--away'' problem, i.e.\ the
vacuum of this system is at $S\rightarrow \infty$. One possible way to avoid it
is to amend the tree level K\"ahler potential  by a non--perturbative
correction, $K=-\ln (S+ \bar S) + \Delta K_\mathrm{np}$. The form of this
correction has been studied in Refs.~\cite{Binetruy:1996xj,Casas:1996zi}. With a
favourable choice of the parameters, the dilaton can be stabilized at a
realistic value $\re S \approx 2$ while breaking supersymmetry,
\begin{equation}
F_S \sim {\Lambda^3 \over M_\mathrm{Pl}} \;.
\end{equation}
The $T$--moduli can be stabilized at the same time by including $T$--dependence
in the superpotential required by $T$--duality \cite{Font:1990nt,Nilles:1990jv}.
In simple examples, the overall $T$--modulus is stabilized at the self--dual
point such that $F_T=0$. This leads to  dilaton dominated supersymmetry
breaking. For $\Lambda \sim 10^{13}$ GeV, the gravitino mass  lies in the TeV
range which is favoured by phenomenology. SUSY breaking is communicated to the
observable sector by gravity \cite{Nilles:1982ik}.

Similar considerations apply to generic models where
  the scale of supersymmetry breaking is
  generated by dimensional transmutation via
  gaugino condensation,
  irrespective of the dilaton stabilization mechanism.

\begin{figure}[!h!]
\centerline{\includegraphics{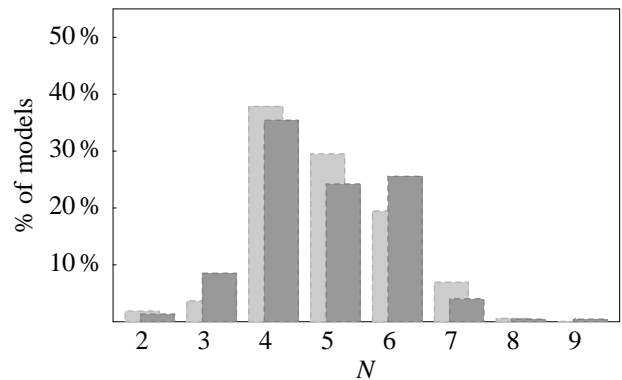}}
\caption{Number of models vs.\ 
the size of largest gauge group in the hidden sector. 
$N$ labels  $\SU{N}$, $\SO{2N}$, $\E{N}$ groups.
The background corresponds to Step 2, while the foreground
corresponds to Step 6.
\label{fig:histogram1}}
\end{figure}

\begin{figure}[!h!]
\centerline{\includegraphics{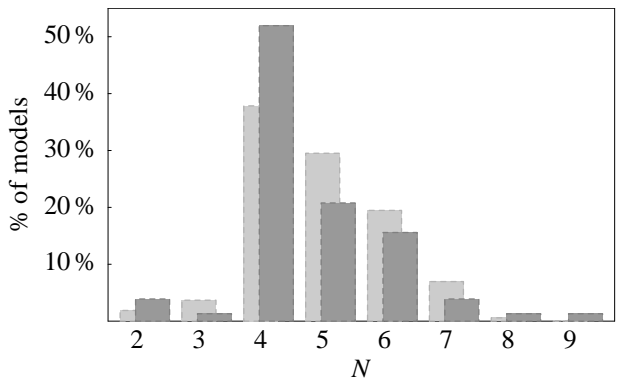}}
\caption{As in Fig.~\ref{fig:histogram1} but with 
models of Step 8 in the foreground.
 \label{fig:histogram2}}
\end{figure}

In Fig.~\ref{fig:histogram1}, we display the frequency of occurrence 
of various gauge groups in the hidden sector (see \cite{Dienes:2006ut} for a related study). 
The preferred size ($N$)  of the gauge groups depends on the conditions
imposed on the spectrum. When all inequivalent models with 2 Wilson
lines are considered, $N=4,5,6$ appear with similar likelihood and
$N=4$ is somewhat preferred. If we require the massless spectrum 
to be the MSSM + vector--like matter, the fractions of models
with $N=4,5,6$ become even closer. 
However, if we further require a heavy top quark and 
the decoupling of exotics at order 8, $N=4$ is clearly preferred
(Fig.~\ref{fig:histogram2}). In this case, $\SU{4}$ and $\SO{8}$ 
groups provide the dominant contribution.
Since all or almost all matter charged under
these groups is decoupled, this leads to gaugino condensation at an 
intermediate scale. 
(We note that before Step 8, gaugino condensation does not
occur in many cases due to the presence of hidden sector matter.) 

\begin{figure}[!h!]
\centerline{\includegraphics{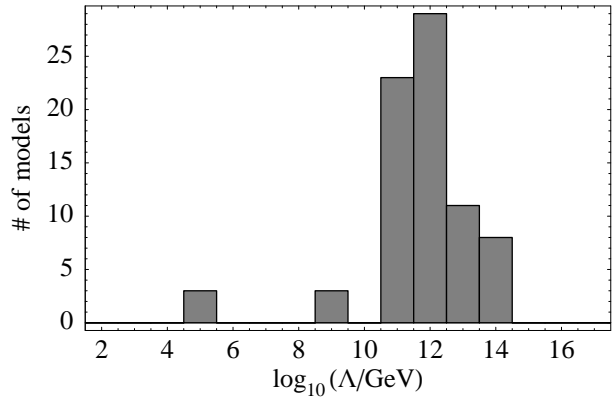}}
\caption{Number of models vs.\ scale of gaugino
condensation. \label{gc}}
\end{figure}

Possible scales of gaugino condensation are shown in 
Fig.~\ref{gc}. These are obtained from Eq.~(\ref{Lambda})
by computing the beta--functions for each case
and using  $g^2(M_\mathrm{GUT}) \simeq 1/2$.

The correlation between the observable and hidden sectors is a result of 
the fact that modular invariance constrains the gauge shifts and Wilson lines 
in the two sectors.  
Moreover, the gauge shifts and Wilson lines determine the massless spectrum via 
the masslessness equations and the GSO projection.

We see that among the promising models, intermediate scale supersymmetry breaking
is preferred. The underlying reason is that realistic spectra require
complicated Wilson lines, which break the hidden sector gauge group.
The surviving gauge factors are not too big (unlike in Calabi--Yau
compactifications with the standard embedding), nor too small.

There are significant uncertainties in the estimation of the supersymmetry breaking scale.
First, the identification of $\langle \lambda \lambda \rangle^{1/3} $
with the RG invariant scale  is not precise. A factor of a few uncertainty in this relation
leads to 2 orders of magnitude uncertainty in $m_{3/2}$.
Also, there could be significant
string threshold corrections   which can affect the estimate. 
Thus, the resulting ``prediction'' for the superpartner masses should be understood
within 2-3 orders of magnitude.

To conclude, we have considered a class of \Z6-II orbifolds with 2 Wilson lines and \SO{10}
and \E6 local GUT structures. The choice of 2 Wilson lines is motivated by the
apparent similarity of the first two fermion generations, while the local GUT
structures are motivated by the quantum numbers of the SM families. We have
found that requiring realistic features in this set of models is correlated with
the supersymmetry breaking scale  such that, in the context of gaugino
condensation, low energy supersymmetry is favoured.   

It would be interesting to extend  these results to Calabi--Yau compactifications
of the heterotic string which also produce promising models
\cite{Braun:2005nv,Bouchard:2005ag}.\\[0.2cm]
\textbf{Acknowledgments.} 
We acknowledge correspondence with K. Dienes. O.L.\ is grateful to B.~Allanach
and B.~Bajc for stimulating discussions.  S.R.\ and A.W.\ would like to
acknowledge research supported in part by the Department of Energy under Grant
No.\ DOE/ER/01545-871 and by the National Science Foundation under Grant No.\ 
PHY99-07949. This work was partially supported by the European Union 6th
Framework Program \mbox{MRTN-CT-2004-503369} ``Quest for Unification'' and
\mbox{MRTN-CT-2004-005104} ``ForcesUniverse''. We would like to thank the Ohio 
Supercomputer Center for allowing us to   use  their resources.


\providecommand{\bysame}{\leavevmode\hbox to3em{\hrulefill}\thinspace}

\end{document}